\RequirePackage{fix-cm}
\documentclass[smallextended]{svjour3}
\smartqed
\usepackage{amsmath,amssymb,paralist,subfigure,graphicx,cite}
\usepackage{algorithmic}
\usepackage{color}

\def\mc{\mathcal}

\newcommand{\ab}[1]{\textcolor{blue}{{#1}}}

\begin{document}
	
	\title{Network-based Control of Epidemic via Flattening the Infection Curve:   High-Clustered vs. Low-Clustered Social Networks 
	}
	\author{Mohammadreza Doostmohammadian        \and
		Hamid R. Rabiee
	}
	
	\institute{ M.  Doostmohammadian \at
		Faculty of Mechanical Engineering, Semnan University, Semnan, Iran, and School of Electrical Engineering, Aalto University, Espoo, Finland. \\
		Tel.: +98-23-31533429\\
		\email{doost@semnan.ac.ir}
		\and
		H. R. Rabiee \at
		ICT Innovation Center, Advanced Information and Communication Technology Research Center, Department of Computer Engineering, Sharif University of Technology, Tehran, Iran. \\ 
		\email{rabiee@sharif.edu}
		}
	
	\date{Received: date / Accepted: date}

	\maketitle
	
	\begin{abstract}
		Recent studies in network science and control have shown a meaningful relationship between the
		epidemic processes (e.g., COVID-19 spread) and some network properties. This paper studies how such network properties, namely clustering coefficient and centrality measures (or node influence metrics), affect the spread of viruses and the growth of epidemics over scale-free networks. The results can be used to target individuals (the nodes in the network) to  \textit{flatten the infection curve}. This so-called flattening of the infection curve is to reduce the health service costs and burden to the authorities/governments. Our Monte-Carlo simulation results show that clustered networks are, in general, easier to flatten the infection curve, i.e., with the same connectivity and the same number of isolated individuals they result in more flattened curves. Moreover, distance-based centrality measures, which target the nodes based on their average network distance to other nodes (and not the node degrees), are better choices for targeting individuals for isolation/vaccination.     
		
	\end{abstract}
	
	\section{Introduction} \label{sec_intro}
	
	Social distancing and individual isolation as the main (non-pharmaceutical)  strategy reduce the spread of epidemics in human interaction networks \cite{block2020social}. On the other hand, vaccination is known to be the key pharmaceutical intervention method to mitigate such pandemics \cite{block2020social}. These strategies are proven to be effective in controlling the recent COVID-19 pandemic. Lockdowns and quarantines are adopted in many countries to reduce the spread of COVID-19, while, more recently, targeted vaccination has significantly lowered the number of infections and severe cases in some countries. One objective of such prevention methods is to reduce the burden on the healthcare system by decreasing the number of COVID-19 cases needed to be hospitalized. This helps to maintain the functioning of the healthcare systems via the so-called \textit{flattening the infection curve} of COVID-19 \cite{block2020social}.
	
	Recently, graph-based strategies have been widely used to control spread of information (or rumor), \cite{haddadi2012spread} and contagious disease \cite{SNAM20} over social networks. Although for different purposes, the spreading procedure seems to be similar over the social network, since both spread processes are closely tied to the structure of the network topology. On the other hand, some compartmental models formulate the epidemic processes based on Markov chain \cite{nowzari2016analysis,SNAM20,ogura2017optimal,chen2017semidefinite}, for example, SIS, SIR, SEIS, SEIR models among others. In this work, we are focused on the first case, i.e., the network structure, to prevent the disease spread over the network. This is typically done by either node removal (isolation) or link removal \cite{pirani_eigvalue}.
	In this work, we use centrality measures and node influence metrics to find and isolate the key nodes in the social network for flattening the infection curve. Such isolation of a node in the network implies either (i) quarantine or (ii) vaccination of that individual in the real-world, where in the latter case we assume that the vaccinated individual may not get re-infected and/or silently transmit the disease as a carrier. In this direction, the outcomes of this work are two-fold:  How targeted (i) vaccination, and/or  (ii)  social distancing may flatten the infection curve. In this direction, some topological properties of the underlying network of individuals may affect the targeted isolation process and spread of epidemics.
	
	Many graph properties of the social network, e.g., clustering, are known to affect even its control features. For example, the (structural)  controllability and observability of complex networks \cite{csl2020,me_complex} and control of SIS epidemic process over the network \cite{SNAM20} can be managed by the clustering and centrality coefficients among other network characteristics. In this work, we investigate how these \textit{structural} control measures can be used to flatten the infection curve and prevent the spread of the epidemic via simulation over some synthetic scale-free networks. These graph topology attributes can help to understand the spreading pattern over the social network and, more importantly, to target the individuals for vaccination (and/or isolation). This is the main idea in this work and our results show meaningful relations between such network properties and the infection curve flattening below some thresholds, where the threshold represents the healthcare system capacity.

	\section{Preliminaries and Problem Setup} 
	\subsection{Social Network Models, Clustering, and Gamma Distribution}
	Consider a human interation network (or the social network) as a graph in which nodes are individuals and links represent acquaintance relationships between those individuals.
	The most common network model capturing the structure of social networks is the scale-free (SF) network, addressing specific empirical social features such as (i) power-law degree distribution, and (ii) preferential-attachment (PA). The first feature implies that in a social network few nodes (individuals), called \textit{network hubs}, have considerably more connections than others. Therefore, the node \textit{degrees}, as the number of friends/neighbours/acquaintances, follow a heavy-tailed (or power-law) distribution, hence the name ``scale-free`` \cite{barabasi2003scale}. This simply implies that the vast majority of individuals in a scale-free social network have very few social connections, while a few important individuals (e.g., celebrities) have plenty of social connections. The latter PA feature in SF networks implies that an individual prefers to socially connect (or attach) to the network hubs more than others.
	
	The two most famous algorithms to generate such SF networks are (i)  Barab\'{a}si-Albert (BA) \cite{barabasi_albert1999}, and (ii) Holme-Kim (HK) \cite{Holme2002clusteringScaleFree} model. The iterative growing procedure in both models are similar in terms of PA rule while differing in terms of \textit{clustering}; a new node $j$ is added to an existent network and makes $m$ new links to the previous set of nodes, where the probability that node $j$ links (attaches) to node $i$ is proportional to its degree $k_i$. The HK algorithm follows a subtle difference; the new node $j$ links to $m_0<m$ existent nodes $i$ and $m_1=m-m_0$  \textit{direct neighbours} of nodes $i$, denoted by $\mc{N}_i$, to make \textit{triads} or \textit{clusters}. Therefore, the latter is also referred to as \textit{clustered scale-free} (CSF) network \cite{Toivonen2006social}. \ab{This triad formation is motivated by \textit{higher clustering in real social networks}; in other words, based on the triad formation two neighboring (friend) nodes most probably have common neighbors (common friends) to represent highly clustered social networks.} 
	
	The motivation behind CSF networks is to better capture the \textit{communities} \cite{mishra2007clustering} and \textit{high clustering} \cite{Toivonen2006social} in social networks. 
	Intuitively, clusters or communities represent groups of individuals which are highly connected internally and sparsely connected externally. In this direction, the CSF model is more clustered than the SF model, implying that it is more probable that two connected individuals (or friends) share more common friends in the CSF model than the SF model. In fact, this shared friendship represents a triad formation in the network and the overall number of such triads directly affects the so-called \textit{clustering coefficient} of the network. In general, the HK network has a larger clustering coefficient than the BA network with the same number of links (i.e., with equal $m$ in the growing procedure). \ab{Recall that it is claimed that the generation of SF networks and the so-called preferential attachment mimics the generation of real social networks. For example, in CSF networks it is more probable that two neighbors (friends) have another common neighbor (a common friend).} 
	
	\ab{Note that other features of the SF and CSF networks are similar and only their clustering coefficient differs. This is the main justification behind using synthetic networks instead of real networks. The existing real datasets and networks in general have different features (e.g., number of nodes, number of links, distribution of the node degrees) other than their clustering. This makes the analysis on real networks less meaningful as we only need to focus on the clustering feature of the networks while keeping the other main features similar. }
	
	Recall that there are two definitions for the global clustering coefficient (GCC). The first one ($\mc{GCC}_1$) is defined as the average of the local clustering coefficient (denoted by $\mc{C}_i$) at all nodes. The local clustering at node $i$ is,
	\begin{align}
		\mc{C}_i = 2\dfrac{ \mbox{Number of links  among nodes in } \mc{N}_i }{  |\mc{N}_i|(|\mc{N}_i|-1) }
	\end{align}
	where the $|\mc{N}_i|(|\mc{N}_i|-1)$ is equal to the total number of possible links among neighbouring nodes in $\mc{N}_i$ (with $|\cdot|$ as the set cardinality). Then, 
	\begin{align}
		\mc{GCC}_1 = \dfrac{\sum_{i=1}^n \mc{C}_i}{n}
	\end{align}
	and the other  one is defined as the probability that two adjacent links make a triad/triangle  by adding a third link (denoted by $\mc{GCC}_2$) \cite{wasserman1994social}, 
	\begin{align}
		\mc{GCC}_2 = 3\dfrac{\mbox{Number of triads}}{\mbox{Number of triplets}}
	\end{align}
    \ab{For the sake of comparison both coefficients can be used, i.e., if $\mc{GCC}_2$ is greater for a sample network $G_1$ as compared with the network $G_2$, then $\mc{GCC}_1$ is also greater. In this work, as it is common in the literature, we use $\mc{GCC}_2$ for comparison and $\mc{GCC}_1$ gives the same comparison perspective.}
    
    It is known that the distance distribution  (also known as the distribution of shortest path lengths) for an SF network follows a semi-Gamma distribution. \cite{steinbock2017distribution,ICNSC,nitzan2016distance}. 
    
\begin{figure}[]
    	\centering
    	\includegraphics[width=3in]{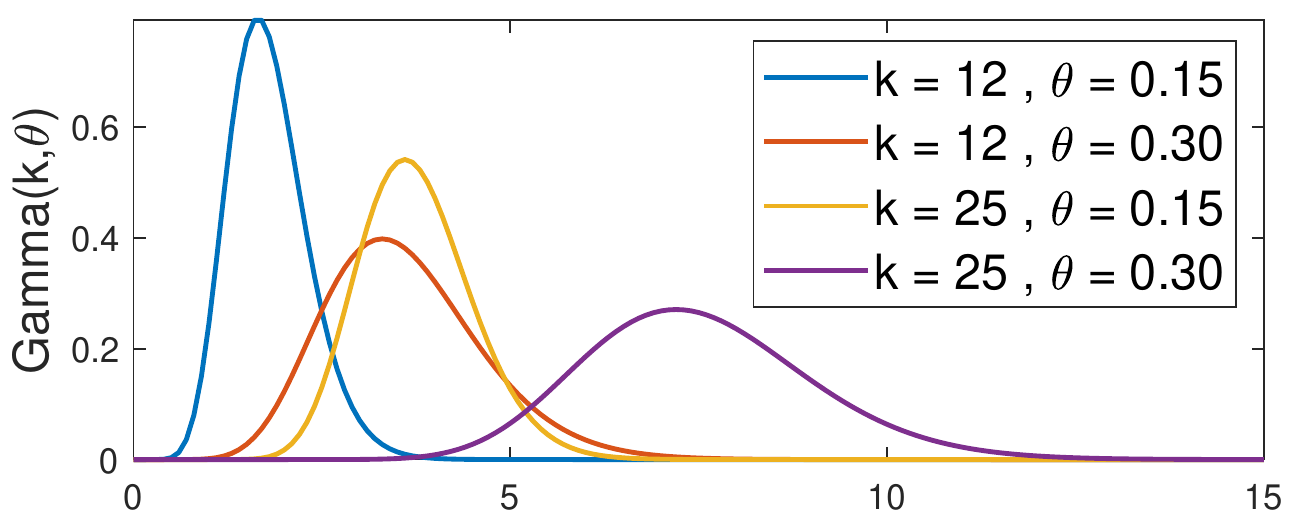}
    	\caption{ Gamma distributions (pdf) for different values of $k$ and $\theta$
    	}
    	\label{fig_gam}
    \end{figure}

    The pdf of Gamma distribution is defined as,
    \begin{align}
    	f(x) = \dfrac{1}{\Gamma(k)\theta^k}x^{k-1}\exp(-\frac{x}{\theta})
    \end{align}
    where $k>0$ is the shape parameter and $\theta>0$ is the scale parameter. Examples of Gamma distributions for different $k,\theta$ are shown in Fig.~\ref{fig_gam}.
    As it is clear from the figure, larger values of $k,\theta$
    imply \textit{flatter} distribution over the range of possible values.

	\subsection{Infection Curve  and Network Distance} \label{sec_infect_curve}
\begin{figure}[tbp]
	\centering
	\includegraphics[width=2in]{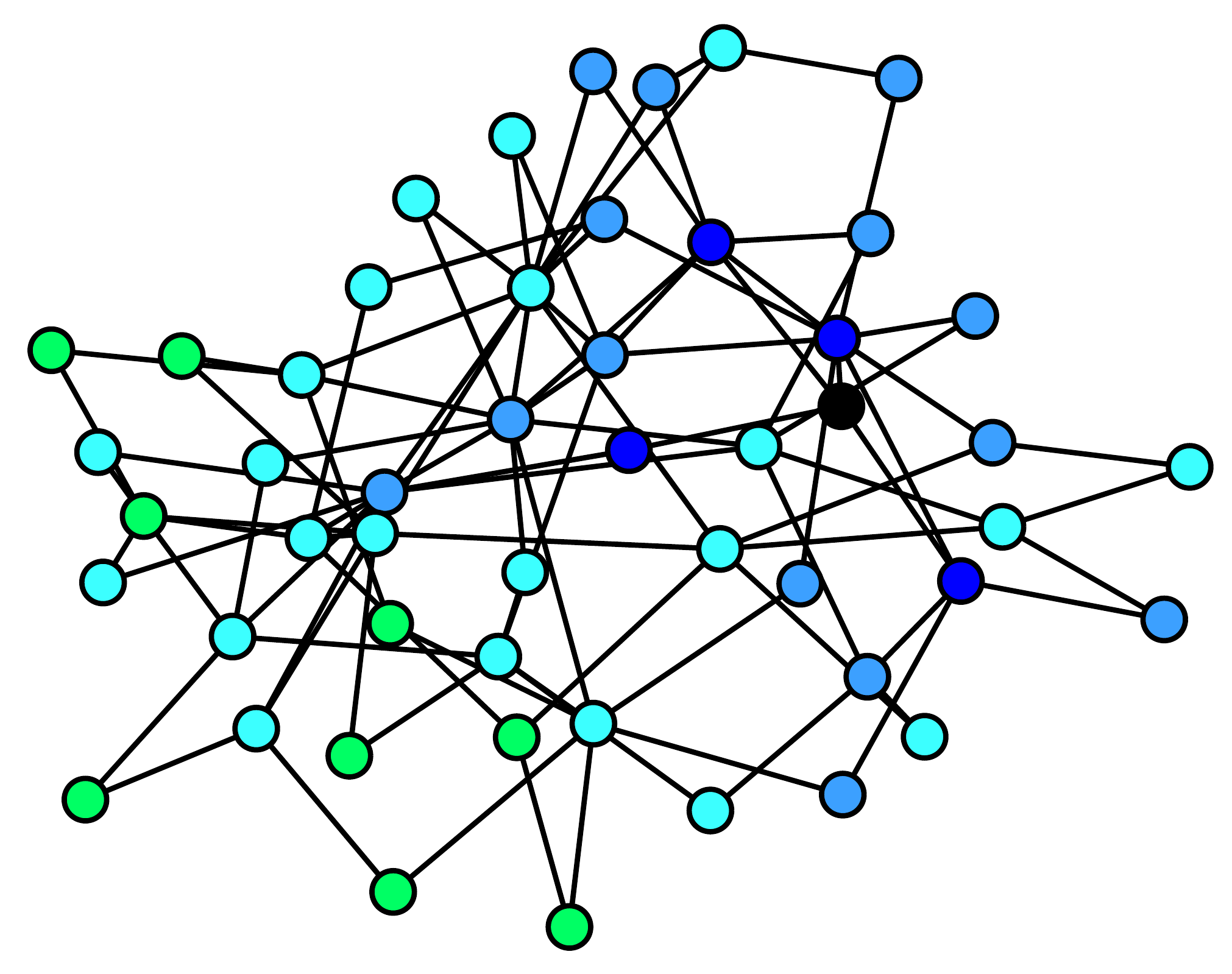}
	\includegraphics[width=2in]{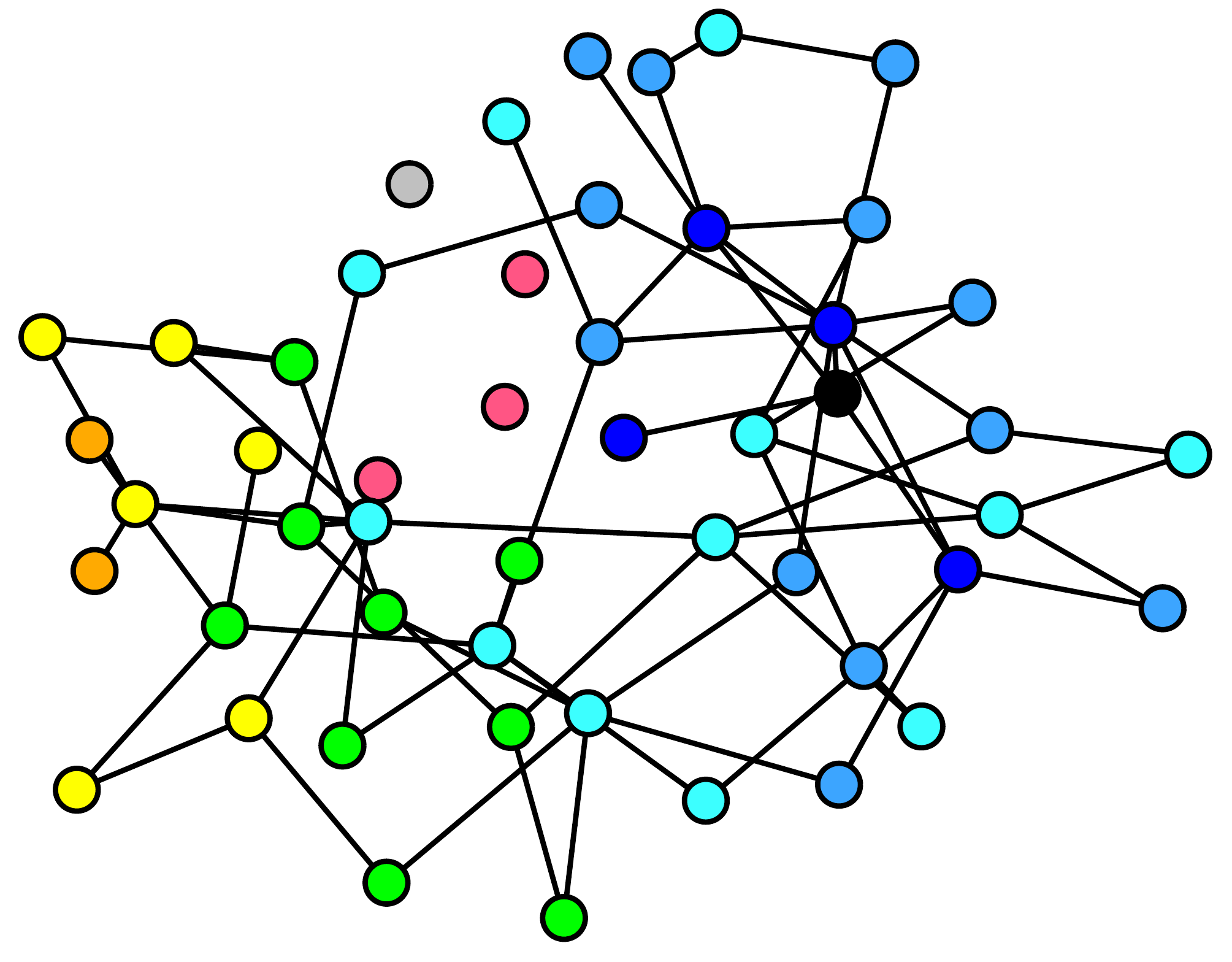}
	\includegraphics[width=2in]{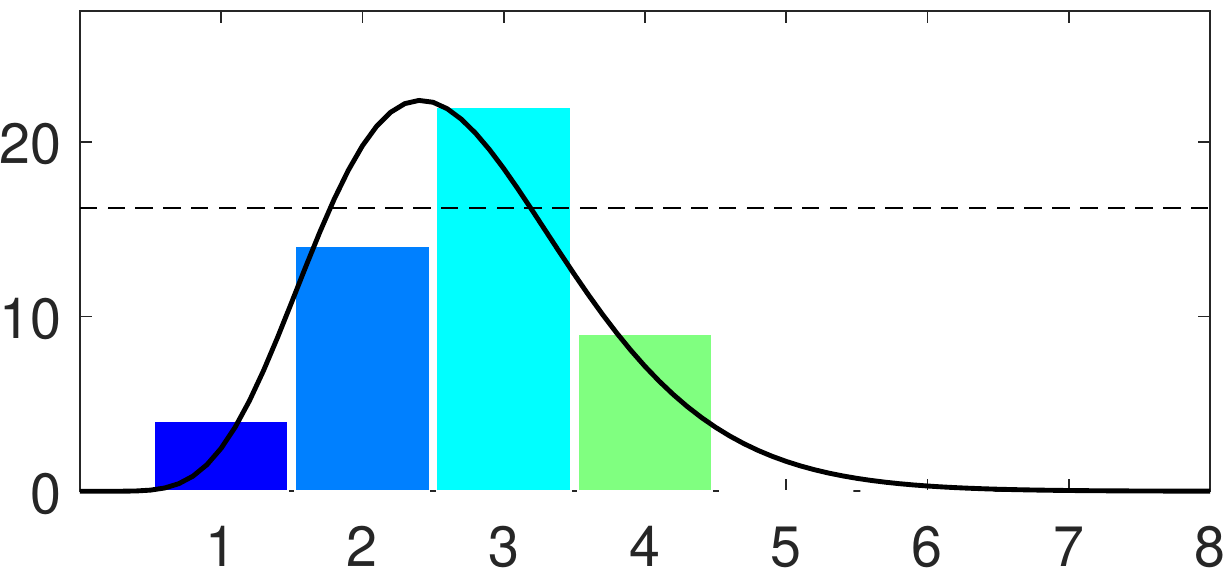}
	\includegraphics[width=2in]{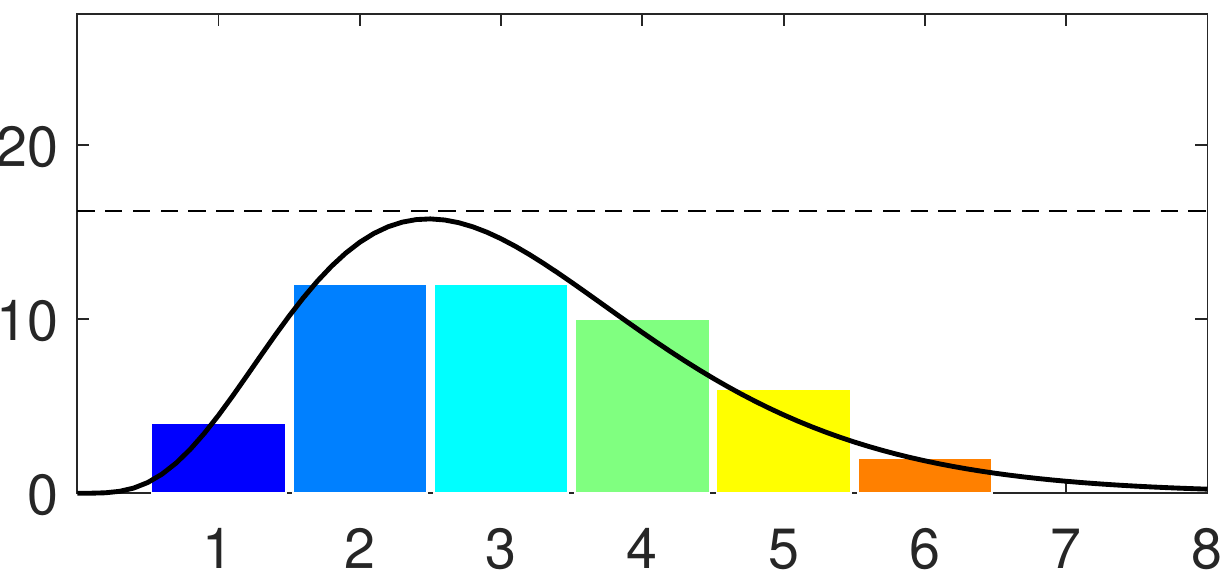}
	\caption{(Left) An example SF network modelling a social network with $50$ nodes as individuals and links as their interactions. The black node represents a hypothetically unknown infected individual (source-node) and at each iteration (a certain amount of time) the disease spreads from infected nodes to their direct neighbours. The nodes of the same colour are at the same distance to the source node, implying that the virus reaches these nodes after the same number of iterations. \ab{In simple words, each bar of the infection curve histogram counts the number of infected nodes at the same distance to the source node.} 
	 The distribution of the distances to the source-node, thus, represents the infection curve over the network which, as shown in the histogram, exceeds the hypothetical capacity of the healthcare system (dashed line). (Right) Isolating specific highly central nodes shown in red colour makes the distances (path lengths) to the (black) source node longer on average. This flattens the infection curve such that it falls below the healthcare capacity.    }
	\label{fig_sample_hist}
\end{figure}

	Define the shortest path length between two nodes $i$ and $j$ as the minimum number of links to reach from node $i$ to node $j$. This is also called the \textit{distance} of the two nodes and, loosely speaking, it takes $d$ steps on the network to connect node $i$ to node $j$ and vice versa. In this direction, node $j$ is sometimes referred to as the \textit{$d$-hop neighbour} (or \textit{distant neighbour}) of node $i$, implying the distance $d>1$ of the two nodes. \ab{In the perspective of epidemics (e.g. the SI model or SIS model with high rate of infection), the distance $d$ in a social network defines the number of infection steps to transfer the disease from the node (individual) $i$ to the node (individual) $j$ and vice versa \cite{block2020social}. This simply means that a chain of \textit{contacts} (represented as links) of at least $d$ infected (or carrier) individuals may transmit the disease between two $d$-hop neighbours.  Therefore, the number of $d$-hop neighbours represents the number of new infections caused by the source node $i$, and the distribution of network distances directly describes the so-called infection curve of social networks. This is better illustrated in the Fig.~\ref{fig_sample_hist}(Right), where the nodes with the same color represent the nodes at the same distance to the infected (black) node. The infection curve is simply defined by the number of same color nodes. Foe example, $9$ green nodes at the distance $d=4$ get infected after $3$ individuals in between every green node and the black node are infected. Counting the number of individuals at distance $d=1,2,3,\dots$ to the infected node makes the infection curve as illustrated in  Fig.~\ref{fig_sample_hist}.}
	
	\subsection{Node Influence Metrics and Centrality Measures} \label{sec_centrality}
	Recall that centrality measures and node influence metrics are quantities that rank the \textit{influence} of different nodes or indicate their \textit{centrality} in the network. Most common such quantities are described here.
	
	\textbf{Degree:} The most well-known centrality measure is the node degree, which represents the number of neighbours (friends) of a node (individual) in the (social) network.
	
	\textbf{Betweenness:} This centrality defines the influence of a node on the flow of entities based on the shortest paths in the network. For every node $i$ its betweenness is defined as the number of shortest paths between different nodes that go through node $i$. A highly-central node, therefore,  bridges the paths between many other nodes.
	
	\textbf{Closeness:} This is another distance-based centrality which defines how close a node $i$ is to the rest of the nodes and is inversely proportional to the average of distances from node $i$ to every other node $j$.
	
	\textbf{Katz:} As a generalization of node degrees, this centrality further counts the number of distant neighbours by a factor $\kappa$. Recall that a distant or $d$-hop neighbour is a node in distance $d>1$ to node $i$. Intuitively, a highly-central node, not only has a high degree itself but is connected to high-degree neighbours which also have their own high-degree neighbours and so on. 
	
	\textbf{PageRank:} Similar to Katz centrality, this measure also relates to node degrees. However, it also evaluates the uniqueness of the links from the neighbours. In other words, a node $i$ has high PageRank if its neighbours $j$ are of high centrality, and further, node $i$ is one of the few neighbours of the nodes $j$.

	\textbf{Expected Force:} This entropy-based epidemiological measure counts the expected number of  \textit{infection force} after two influence transmissions with respect to node degrees.

	In general, finding the centrality of the nodes requires global or local knowledge of the network. Some of these centrality measures can be defined locally \cite{haddadi2013hidden,ICNSC}, while others need information of the entire network topology.

	\subsection{Problem Statement}
	During a pandemic, the healthcare system can break down when the number of severely infected cases exceeds its serving capability. Therefore, epidemiologists propose preventive solutions and mitigation techniques to reduce the infection rate by \textit{flattening the infection curve}, which refers to the strategies to slow down the spread of the epidemic. This helps to keep the peak number of infected individuals in the capacity range of the healthcare systems. This work aims to flatten the curve via targeted isolation strategies by cutting all the links of the isolated node (individual) to its neighbours. This either implies that the targeted individual is in quarantine (with no interaction with others) or vaccinated (is not a carrier node to transmit the virus).  
	
	Recall that for two networks with the same number of links, the network with a longer path length has a flatter infection curve. This means that the nodes (individuals) in one network are more distant due to its particular graph topology while having the same contact prevalence (number of links in the network). In this direction, we first compare the distance distribution in two key social network structures, the SF and CSF models via Monte-Carlo (MC) simulations. This helps to understand the effect of clustering on the infection curve. Second, we study flattening the infection curve via targeted isolation to increase the nodes' distances in the social network (see Fig.~\ref{fig_sample_hist}) via MC simulations. We particularly, choose the  target nodes (individuals)
	based on their centrality or influence in the network, and compare the results for both SF and CSF models. In this direction, we study the parameters of the semi-Gamma function associated with the infection curve after isolating (removing) central nodes in both SF and CSF networks.  This helps to understand the role of different key central nodes for the epidemic spread over the network.
	
    \ab{After target node removal/isolation the network may become islanded and partitioned into components. In our simulation we take care of this and avoid such situations. This is done by considering highly connected networks and isolating fewer nodes to avoid network disconnectivity. If the network changes into many components, one may consider the average distance in each component. In all our simulations in this paper the network remains one connected component while we isolate the central nodes.}

	\section{Centrality-based Infection Curve Flattening}
	
	\subsection{Infection Curve versus Clustering Coefficient}
\begin{figure}[tbp]
		\centering
		\includegraphics[width=1.5in]{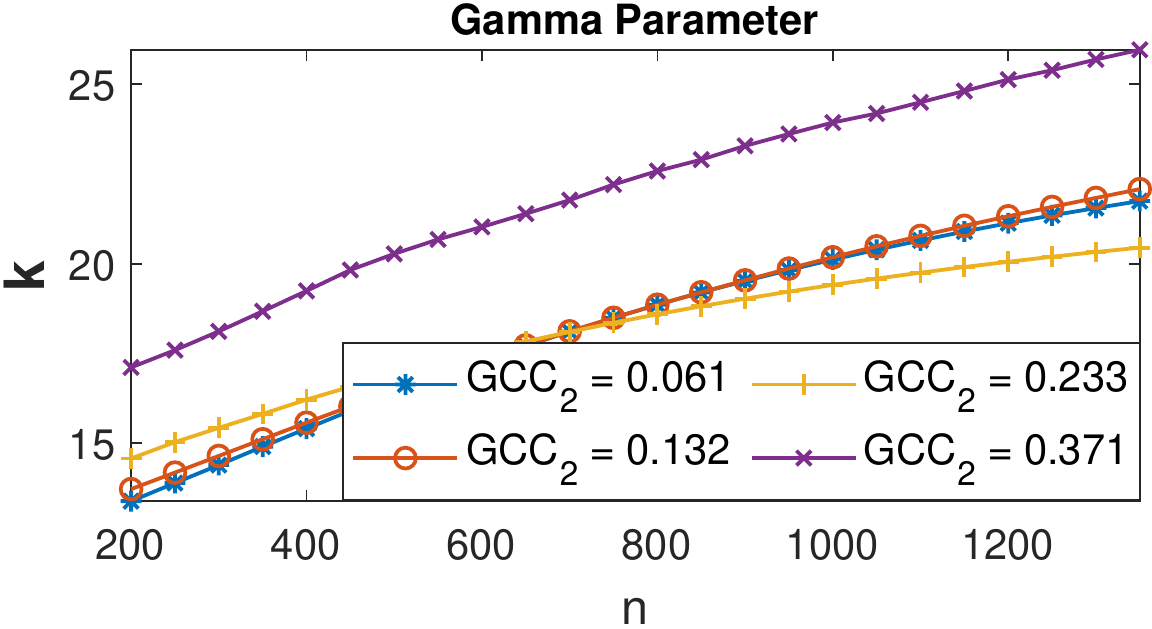}
		\includegraphics[width=1.5in]{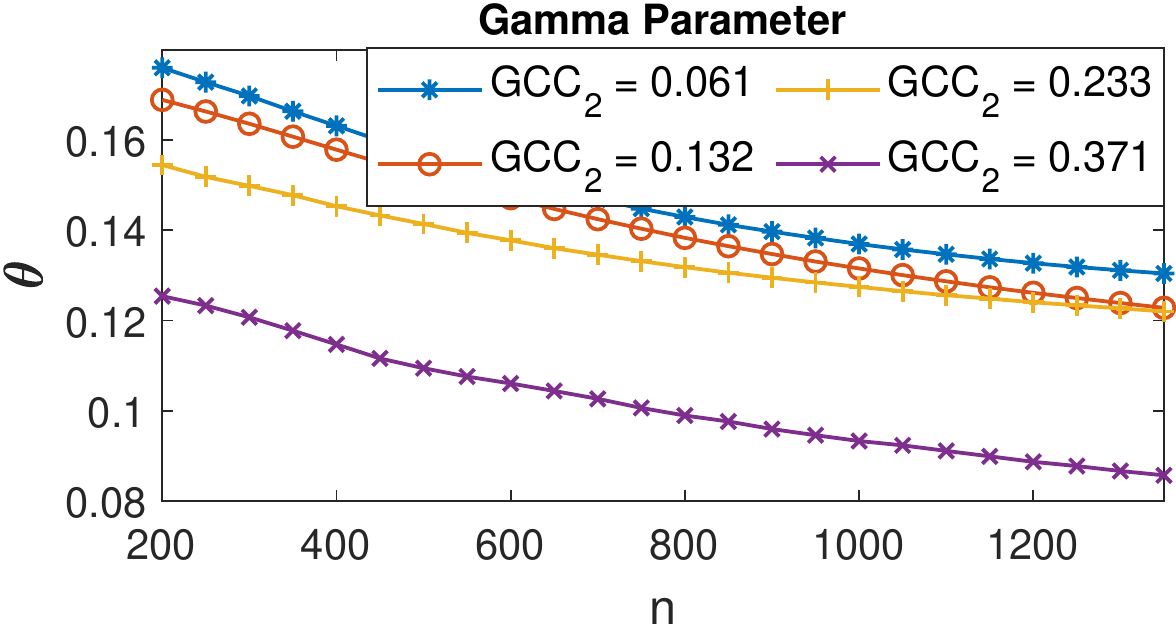}
		\includegraphics[width=1.5in]{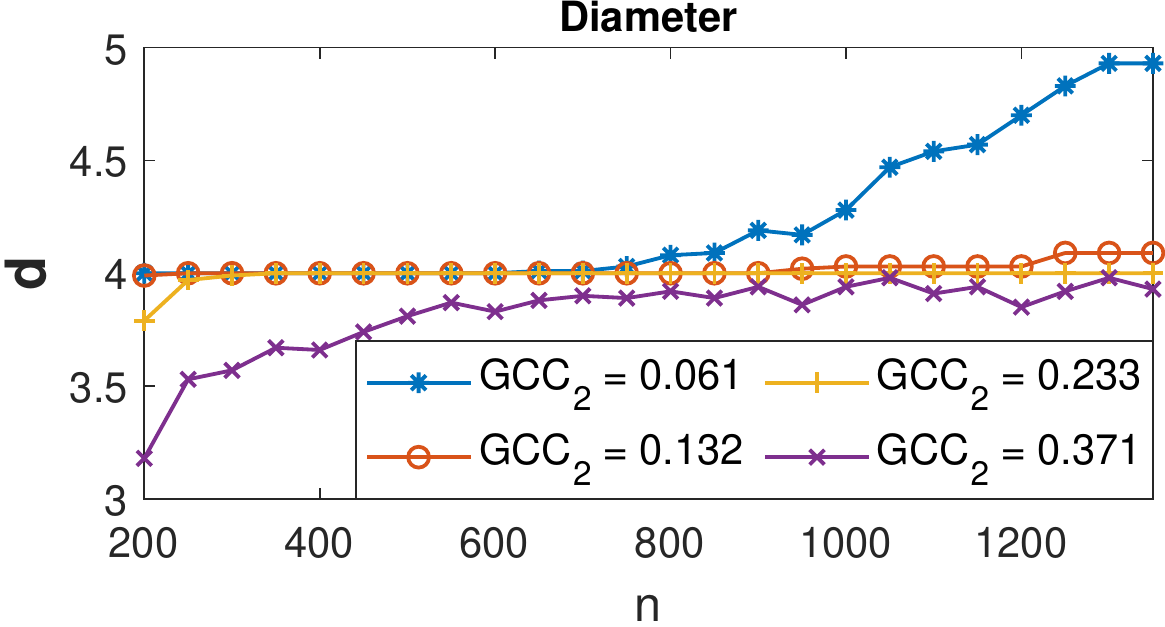} 
		\raggedleft
		\includegraphics[width=1.38in]{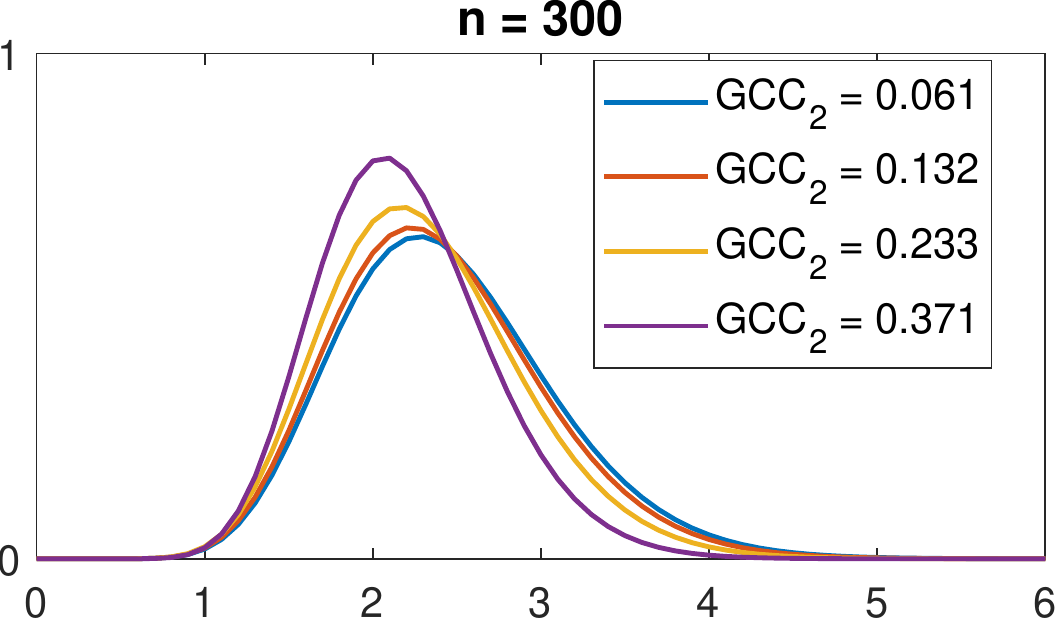} 
		\hspace{0.07in}
		\includegraphics[width=1.38in]{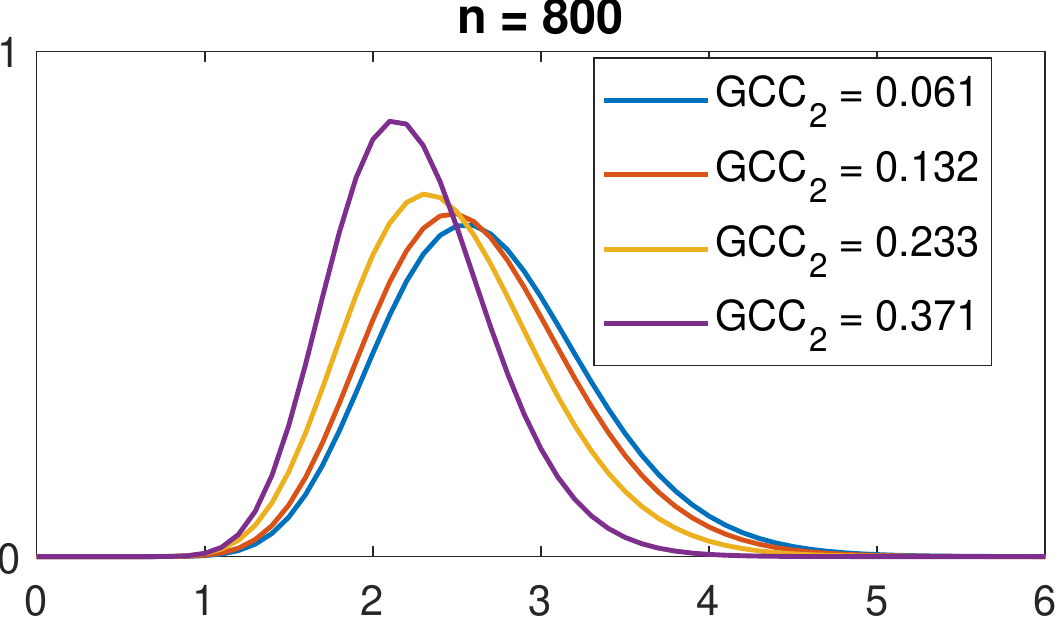}
		\hspace{0.07in}
		\includegraphics[width=1.38in]{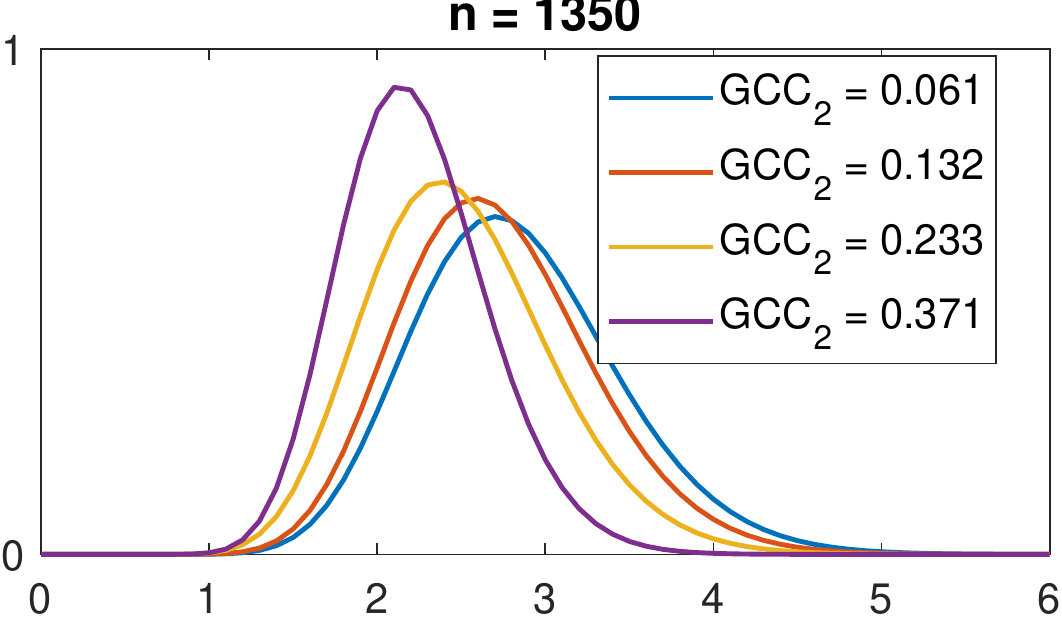}
		\caption{ (Top) Change in the diameter and Gamma parameters of the distance distribution for $3$ HK (with high GCC) and $1$ BA (with low GCC) networks are represented versus different network sizes.  (Bottom) The distance distribution of the HK and BA models for networks of size $n=300,800,1350$. 
		}
		\label{fig_MC_curve}
	\end{figure}

	In this subsection, we first study how network clustering affects the infection curve (distance distributions) in SF networks via empirical MC simulation.  
	We generate different CSF networks by changing the triad formation factor. Following the HK model, we generate networks for different values of $m_0$ (where $m_0 = 0$ represents the BA model). The empirical results are averaged over $100$ MC trials and are shown in Fig.~\ref{fig_MC_curve}. The values of Gamma parameters and network diameter are shown for different network sizes. By increasing $m_0$ (and clustering coefficient) the network diameter is increased, with increasing $K$ parameter and decreasing $\theta$ parameter (in average).
	
	Fig.~\ref{fig_MC_curve} shows how the change in clustering coefficient affects the distance distribution curves. For a larger change in the clustering coefficient, the Gamma parameters may change sharply. Following the examples given in Fig.~\ref{fig_gam} and from the illustrations in Fig.~\ref{fig_MC_curve}, the infection curves for highly clustered networks are taller and narrower as compared to the networks with lower clustering. Recall that in Fig.~\ref{fig_MC_curve}, each graph is given for the same number of links and the same number of nodes, which implies the same connectivity and linking over the social network. \ab{Both definitions of the clustering coefficient can be used for the comparison in this subsection and the next subsection.} It might be thought that it is more challenging to flatten the distance curves associated with highly clustered networks (due to their taller curves).
	However, as we will see in the coming subsections, such networks give flatter outcome curves after isolating the same number of nodes.  
	
	
	\subsection{Monte-Carlo Simulations for Node Isolation}
	In this section, we consider two scenarios to flatten the infection curve by removing some target nodes from the network. The nodes are ranked based on their influence metrics and centralities discussed in Section~\ref{sec_centrality}. The nodes with higher centralities are chosen as the target nodes.
	
\begin{figure}[tbp]
		\centering
		\includegraphics[width=1.15in]{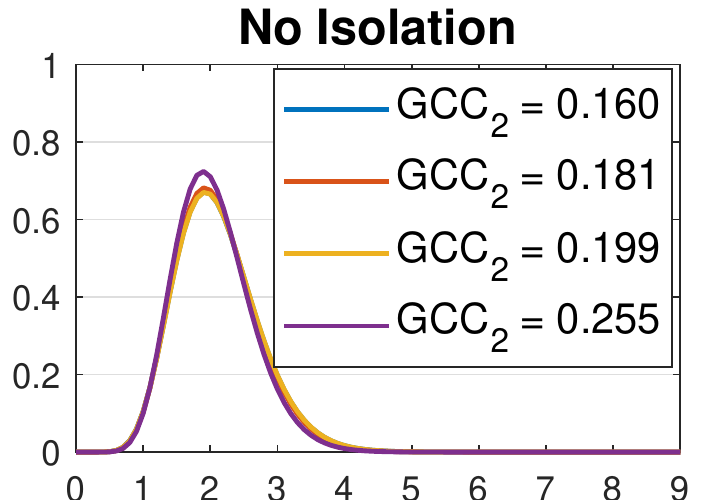}
		\includegraphics[width=1.1in]{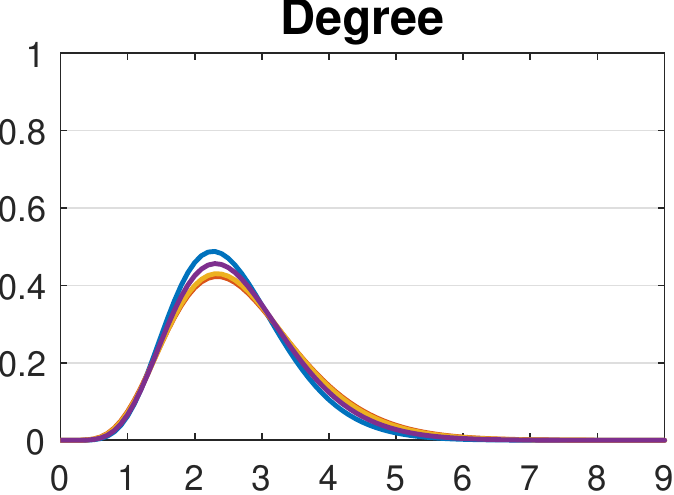}
		\includegraphics[width=1.1in]{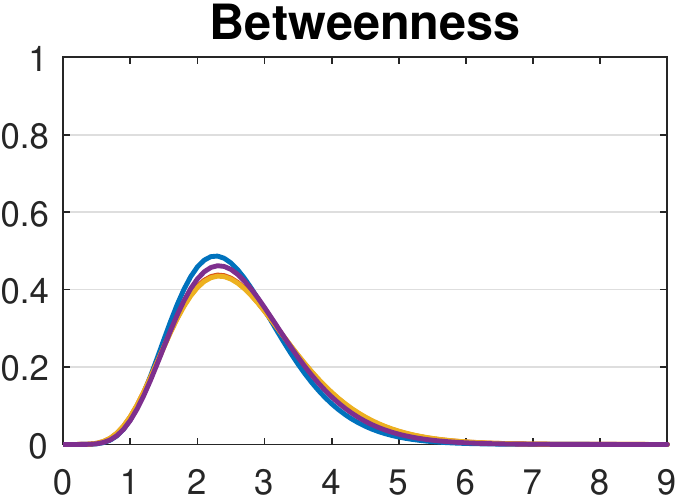}
		\includegraphics[width=1.1in]{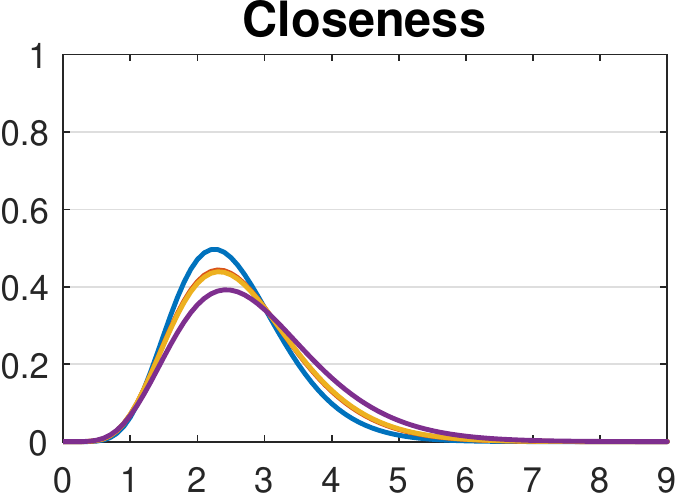} 
		\includegraphics[width=1.1in]{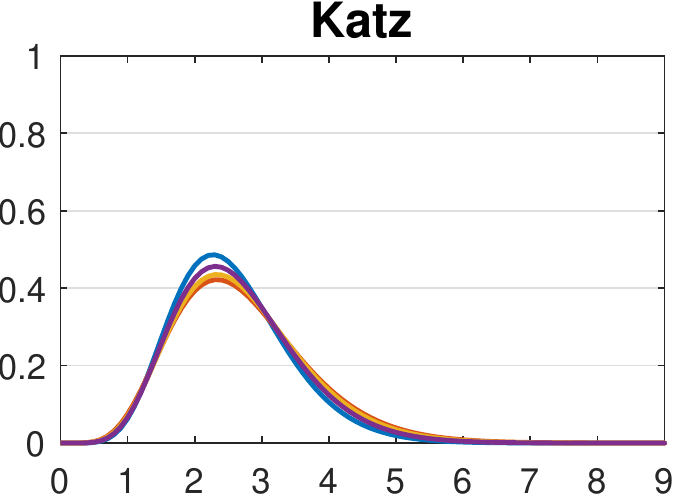}
		\includegraphics[width=1.1in]{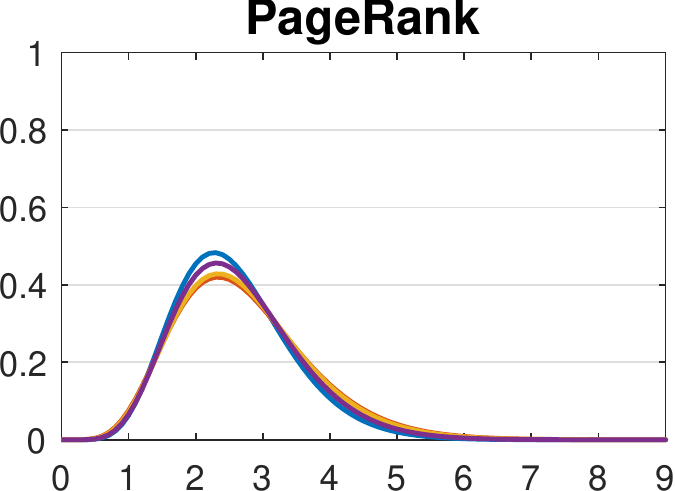}
		\includegraphics[width=1.1in]{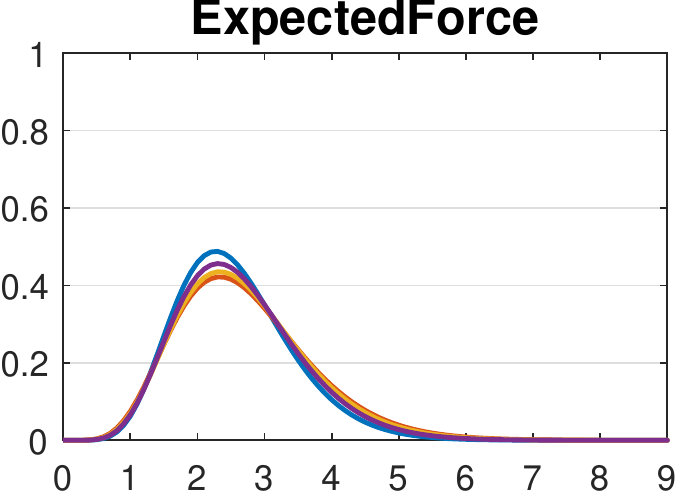}
		\caption{ The infection curve (distance distribution) of $3$  HK networks (with high GCC) and $1$  BA network (with low GCC) before and after isolation of $5\%$ highly central nodes. The simulation is repeated using different centrality measures. }
		\label{fig_removal_5p}
	\end{figure}

	\textbf{Scenario 1:} we isolate/remove $5 \%$ of the community based on their centrality ranking; in other words,  top $5 \%$ nodes with high centrality measures are chosen as the target nodes.  This simulation is done over the SF networks with a different number of triads (and clustering coefficients) of $1000$ nodes.  $50$ highly-central nodes are isolated and the distance distributions of the modified networks (after node isolation) are provided. \ab{We repeat this for $20$ Monte-Carlo trials using the centrality measures in Section~\ref{sec_centrality}, and normalize (take the average of) the distance distribution accordingly. Fig.~\ref{fig_removal_5p} shows the change in the gamma distribution associated with the distances after (target) node isolation. As it is clear from the figures, the gamma distribution are flattened for all networks while the change in highly-clustered networks is more significant. This implies that the targeted node isolation strategy more flattens the highly clustered CSF networks as compared to low-clustered SF networks. }
	
\begin{figure}[tbp]
		\centering
		\includegraphics[width=1.1in]{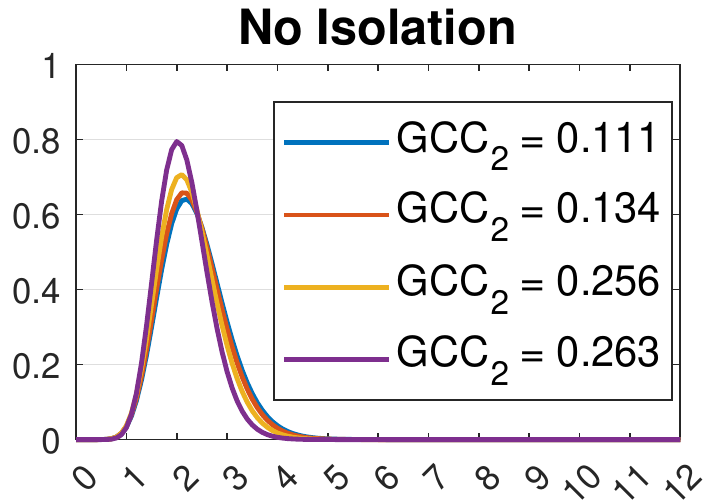}
		\includegraphics[width=1.1in]{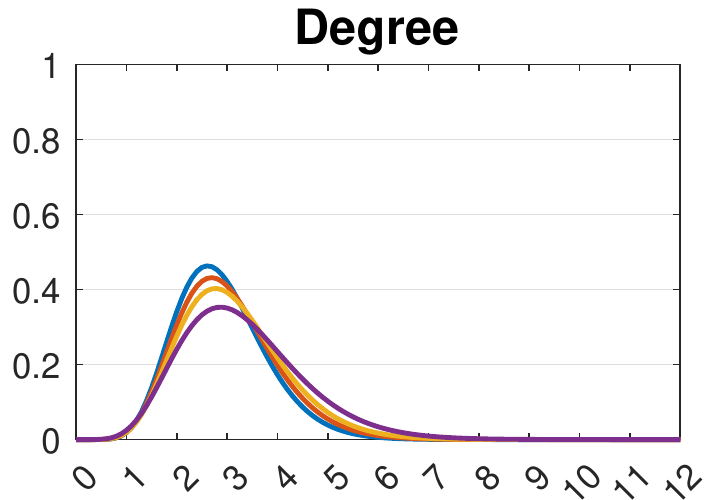}
		\includegraphics[width=1.1in]{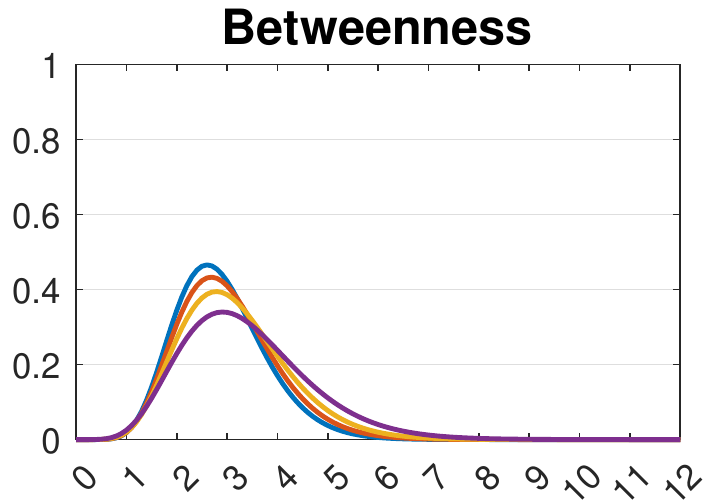}
		\includegraphics[width=1.1in]{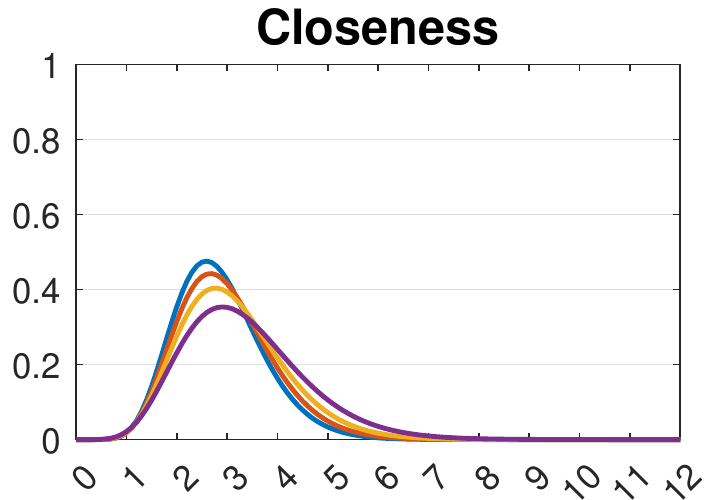} 
		\includegraphics[width=1.1in]{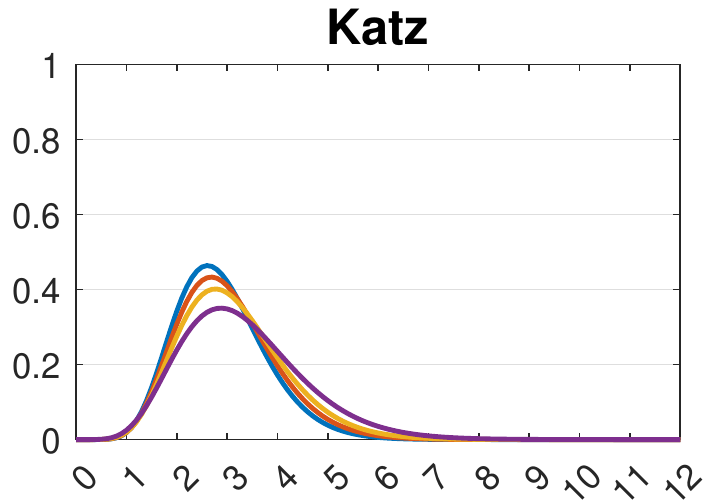}
		\includegraphics[width=1.1in]{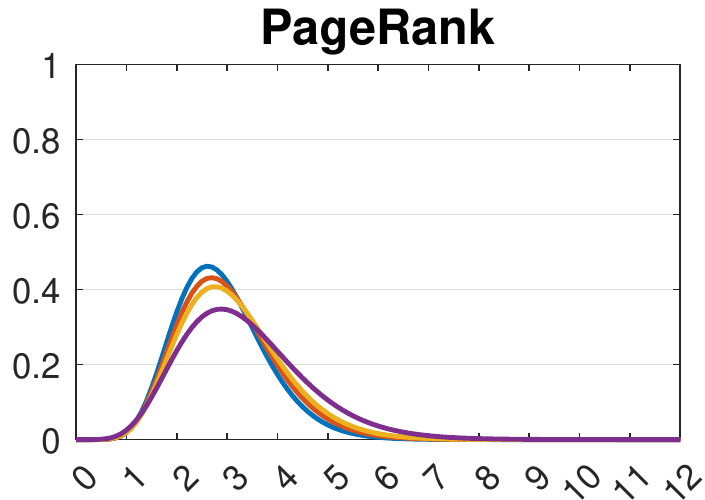}
		\includegraphics[width=1.1in]{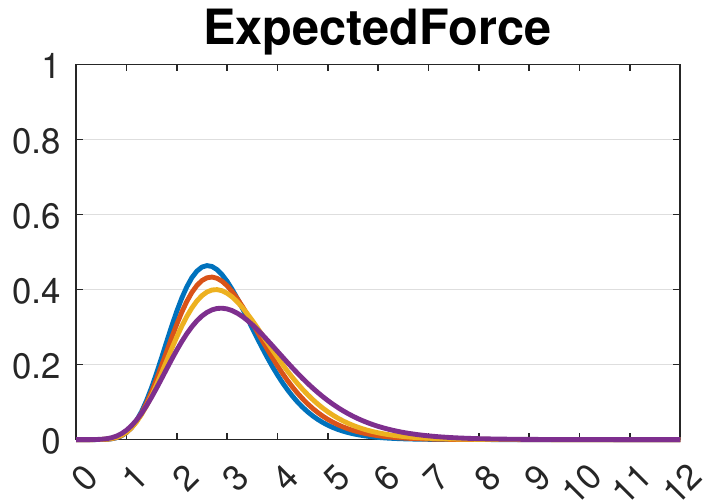}
		\caption{ The infection curve of the BA  and HK (or CSF) networks before and after isolation to flatten the infection curves below some given threshold. This threshold represents healthcare system capacity. The title of the figure represents the measure chosen for the isolation of highly central nodes/individuals.   }
		\label{fig_removal_threshold}
	\end{figure}

	\textbf{Scenario 2:}
	in this simulation, we consider a  
	certain hypothetical capacity for the health-care system (or the \textit{threshold}). \ab{We isolate/remove the highly central nodes (based on different centrality measures) and the number of these target nodes are increased up to the point that the infection curve (distance distribution) goes below the given threshold of $0.5$. We generating different SF and CSF networks of $1000$ nodes with tuned clustering (as in the previous simulation) for $20$ MC simulation trials, and isolate the same number of target nodes at all (highly-clustered and low-clustered) networks. The results are compared based on different centrality measures and clustering coefficient of the networks. As in the previous scenario, the infection curves before and after simulations are shown in the Fig.~\ref{fig_removal_threshold} and clearly the infection curve flattening is more significant for highly clustered networks as compared to low clustered networks. For example the purple tall gamma distribution associated with high $\mc{GCC}$ CSF network is drastically flattened after isolating the same number of nodes (as compared to other three low-clustered networks).}
	
\begin{figure}[tbp]
	\centering
	\includegraphics[width=1.25in]{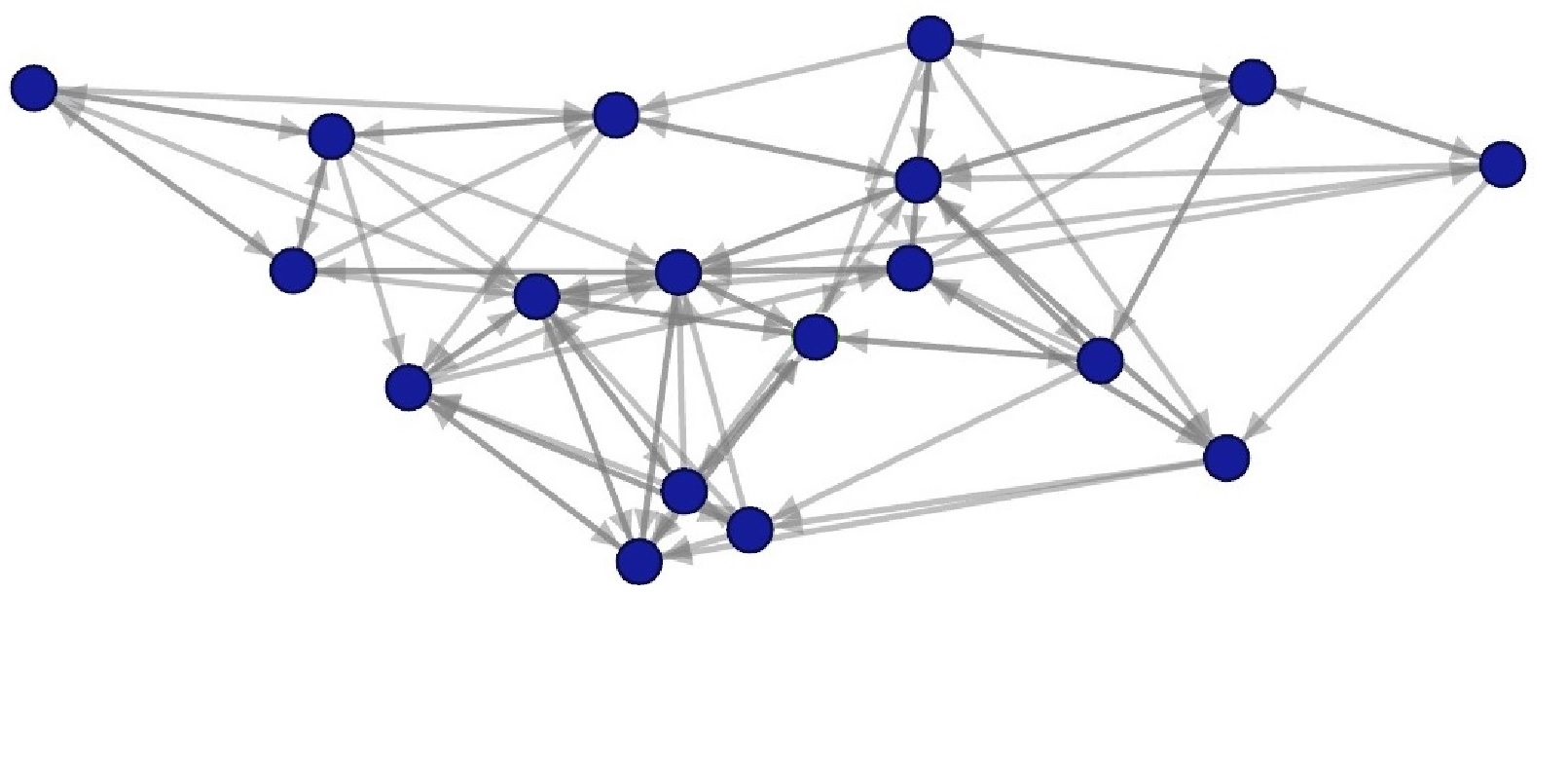}
	\includegraphics[width=1.1in]{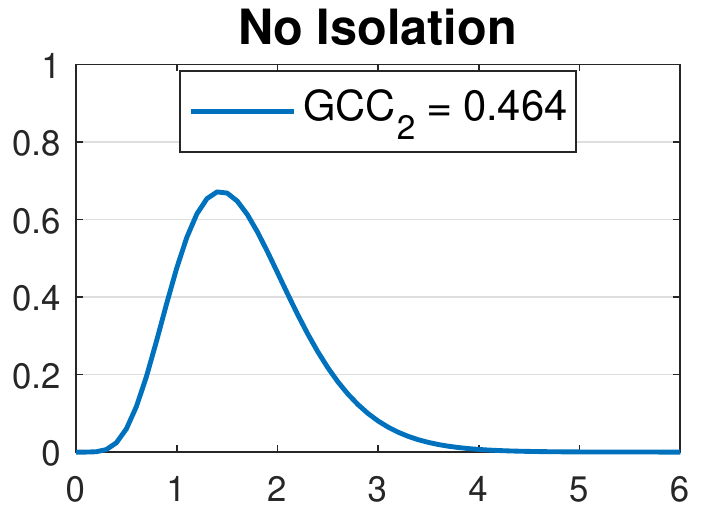}
	\includegraphics[width=1.1in]{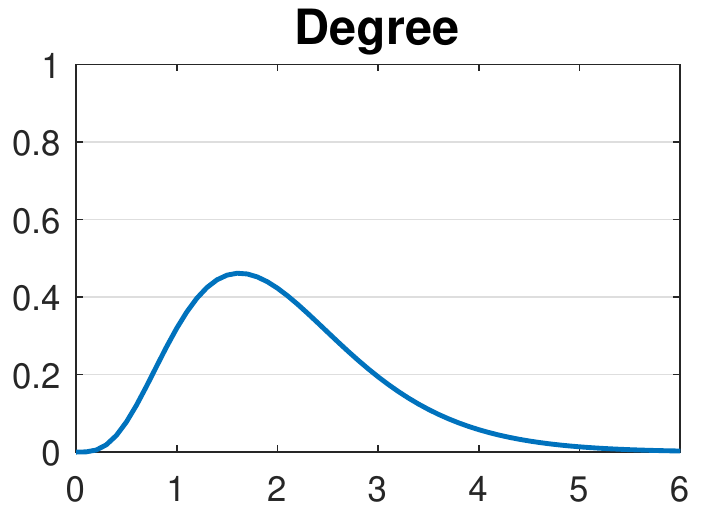}
	\includegraphics[width=1.1in]{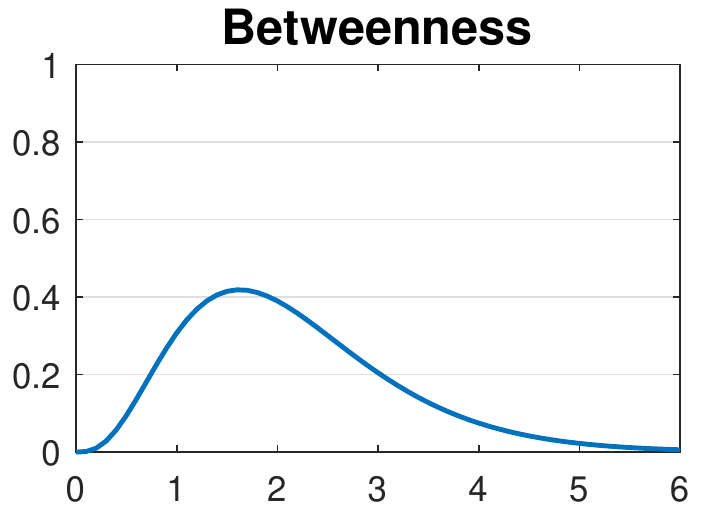}
	\includegraphics[width=1.1in]{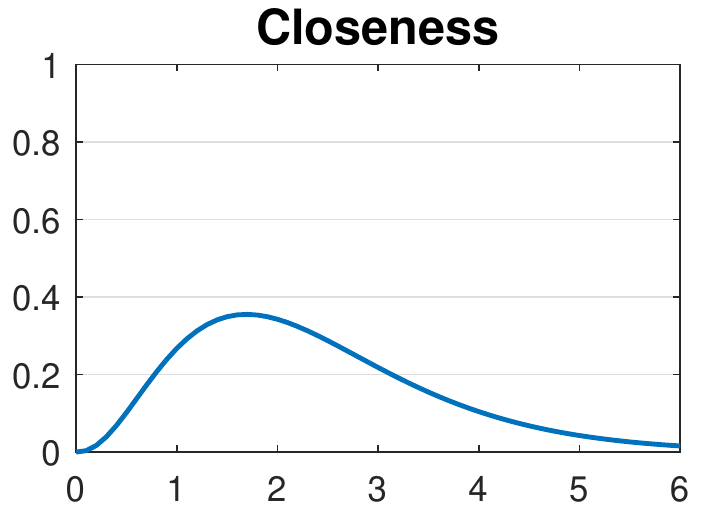} 
	\includegraphics[width=1.1in]{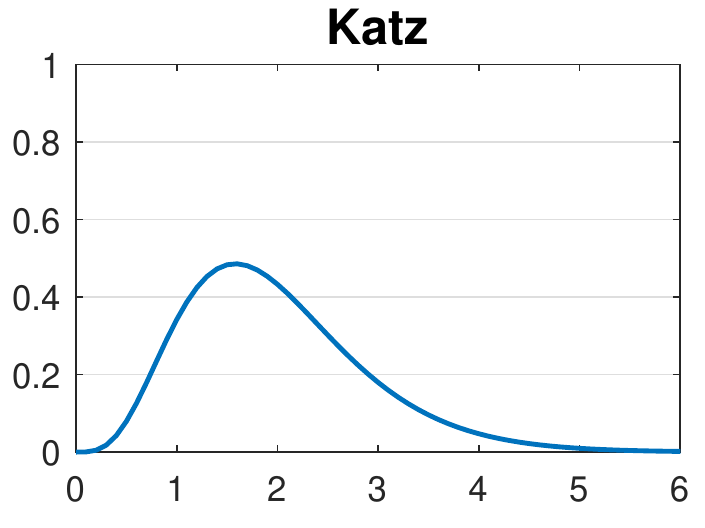}
	\includegraphics[width=1.1in]{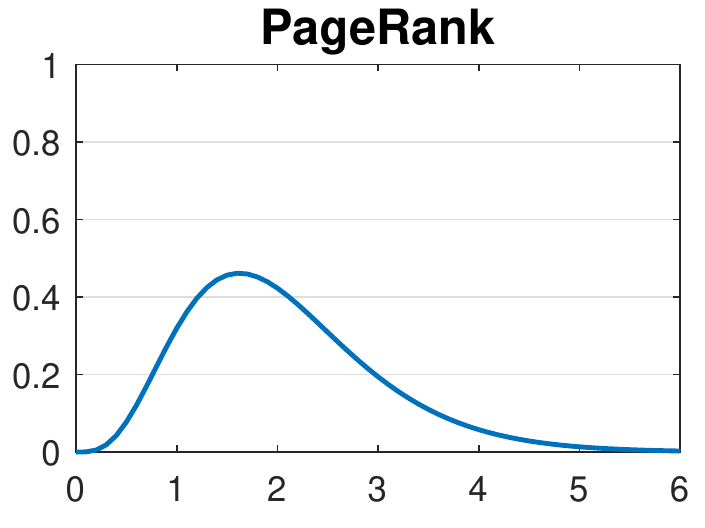}
	\includegraphics[width=1.1in]{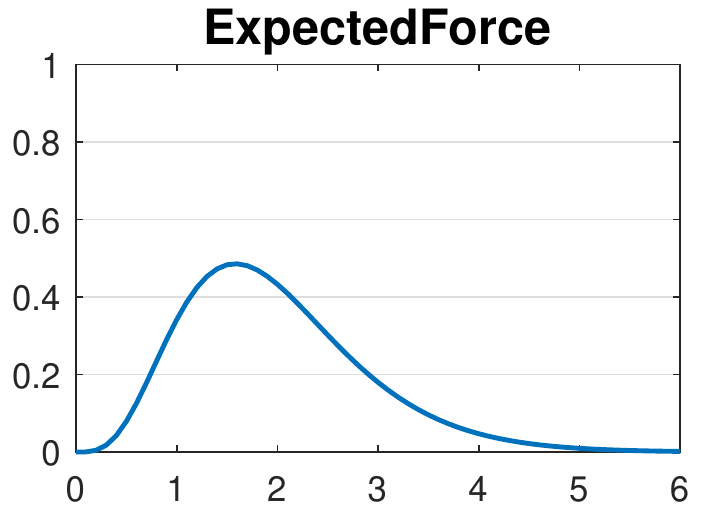}
	\caption{ \ab{The  Sampson's Monastery network and its infection curves before and after isolation of the target central nodes based on different centrality measures.}  }
	\label{fig_removal_real}
\end{figure}

	\ab{\subsection{Simulations over Real Networks}
    A real social network of Sampson's Monastery Data of $18$ individuals (nodes) and $26$ social contacts (links) is considered \cite{uci}. The network represents the social relations among a group of men (novices) who were preparing to join a monastic order. Infection starts from a randomly-chosen node and $4$ central target nodes based on different centrality measures are isolated/removed. The infection curves before and after node isolation are shown in Fig.~\ref{fig_removal_real}. For this example the clustering coefficient is $0.464$. Note that the reason behind using synthetic networks is to tune this clustering coefficient while keeping the same number of nodes and links (and similar power-law degree distribution). This is not an option for real networks because for two different real networks many features other than the $\mc{GCC}$ change which may affect the infection curve flattening.
    }
	
	\subsection{Some Discussions}
	One can observe based on the MC simulations that,  for all centrality-based node isolation cases, the blue curve related to the BA model is over the other three curves related to the HK model. This implies that, in general, the centrality-based node isolation strategy works (slightly) \emph{better} on the \emph{highly-clustered} networks in terms of flattening the infection curve. This is more clear from isolation based on the \emph{closeness} centrality, which shows much better performance in flattening the \emph{highly clustered} HK  networks (in both Fig.~\ref{fig_removal_5p} and Fig.~\ref{fig_removal_threshold}). Recall that closeness is a distance-based centrality, i.e., it relates to the average of the minimum shortest path length to the other nodes, while other centrality measures are mostly degree-based. 
	
	\section{Concluding Remarks}
	Network pruning (with either node or link removal) \cite{csl2020,themis2017loco} can be used to tune properties of the network, e.g., the clustering. In this work, we adopted the idea of network pruning to flatten the node distance distribution of the Scale-Free networks, resembling the idea of flattening the infection curve in social networks. Our comparative simulation results on both CSF and SF networks (resp. HK and BA models) show that the highly clustered CSF networks, although have narrow and tall infection curves, they become flatter after isolating the same number of nodes. For the simulations, we considered the same number of linking and connectivity in both HK and BA models. This implies that clustered social networks, although may show more positive cases in a shorter time period (i.e., narrower infection curves) before preventive measures, they end up with fewer cases over a longer period (flatter infection curves) after isolating few \textit{highly central individuals}. Our results also show that these preventive measures even work better on CSF networks if the central nodes for isolation are chosen based on their network distance to other nodes rather than their node degrees. These results may help the idea of flattening the curve to spreading the number of new cases over a longer period so that more people have better access to healthcare services.
	
	\textcolor{blue}{As future research simulating similar strategies over Small-World (SW) networks  \cite{watts1998smallworld} is a promising research direction. New strategies should be developed to tune the clustering coefficient while keeping the same number of links and network connectivity. }
	
	\section*{Acknowledgements}
	The calculations and simulations presented in this paper are done over Triton cluster, using computer resources within the Aalto University School of Science “Science-IT” project. 
	The first author would like to thank Hossein Vahid Dastjerdi for his help with simulations over Triton cluster, and to thank Usman A. Khan and  Themistoklis Charalambous for their helps and comments. The authors acknowledge the use of some MATLAB codes from Koblenz Network Collection (KONECT) \cite{konect}.

	\bibliographystyle{spphys} 
	\bibliography{bibliography}
	
\end{document}